\documentclass[twocolumn]{aastex62}

\graphicspath{{./}{figures/}}

\submitjournal{ApJ}

\shorttitle{Electron-only reconnection}
\shortauthors{Vega et al.}

\begin{document}

\title{Electron-only reconnection in kinetic Alfv\'en turbulence}

\correspondingauthor{Cristian Vega}
\email{csvega@wisc.edu}

\author{Cristian Vega}
\affiliation{Department of Physics, University of Wisconsin-Madison, Madison, WI 53706, USA}

\author{Vadim Roytershteyn}
\affiliation{Space Science Institute, Boulder, Colorado 80301, USA}

\author{Gian Luca Delzanno}
\affiliation{T-5 Applied Mathematics and Plasma Physics Group, Los Alamos National Laboratory, Los Alamos, NM 87545, USA}

\author[0000-0001-6252-5169]{Stanislav Boldyrev}
\affiliation{Department of Physics, University of Wisconsin-Madison, Madison, WI 53706, USA}
\affiliation{Space Science Institute, Boulder, Colorado 80301, USA}



\begin{abstract}
We study numerically small-scale reconnection events in kinetic, low-frequency, quasi-2D turbulence (termed kinetic-Alfv\'en turbulence). Using 2D particle-in-cell simulations, we demonstrate that such turbulence generates reconnection structures where the electron dynamics do not couple to the ions, similarly to the electron-only reconnection events recently detected in the Earth's magnetosheath by \citet{phan2018}. Electron-only reconnection is thus an inherent property of kinetic-Alfv\'en turbulence, where the electron current sheets have limited anisotropy and, as a result, their sizes are smaller than the ion inertial scale.  The reconnection rate of such electron-only events is found to be close to~$0.1$.
\end{abstract}



\section{Introduction} 

Magnetic reconnection has long been considered as one of the fundamental mechanisms of plasma heating and particle acceleration in astrophysical and space plasmas \cite[e.g.,][]{biskamp_magnetic_2005,priest2007}. Analytical and numerical studies have established that in a collisionless, two-component plasma, where the ions and electrons are not strongly coupled, a reconnection layer generally has at least a two-scale structure: inside an ion-scale current sheet whose thickness is comparable to the ion inertial length $d_i$, there is a thinner electron-scale current sheet, whose thickness in on the order of the electron inertial length $d_e$~\cite[e.g.,][]{shay2007,karimabadi2007}. Magnetic-field lines, frozen into the electrons, reconnect at the electron scale on a short electron time scale, setting up electron outflows along the current sheet. At much larger scales, of the order of $d_i$ and sometimes significantly exceeding it, the electron outflows  couple to the ions, leading to the plasma outflow from the reconnection layer on the ion scales. 

Recent observations of the Earth's magnetosheath by the NASA Magnetospheric Multiscale (MMS)  mission have,  however, drawn attention to reconnection events where only the electron reconnection layers have been detected, without the accompanying ion ouflows \cite{phan2018}, see also previous studies  \cite{yordanova2016,voros2017,wilder2018}. Similar electron-only reconnection events were found in the hybrid Vlasov-Maxwell simulations in \cite{califano2018} when fluctuation energy was injected close to the ion kinetic scale. In these electron-only reconnection events, typical signatures of reconnection where identified, such as the characteristic magnetic and current profiles, the presence of diverging bi-directional electron-Alfvénic  jets, but no ion jets were found. An explanation suggested by \cite{phan2018} pointed out  that the ion dynamics was not observed because the overall dimensions of the sheets were smaller than the typical scale (several $d_i$) required for efficient ion coupling to the electron streams. This suggestion is consistent with the results of previous hybrid simulations \citep[e.g.][]{mandt1994} and PIC simulations \citep[e.g.][]{pyakurel2019}, that indicate that a length of the reconnection layer of about $5-10d_i$ is needed to observe a ``traditional'' reconnection site, with  efficient coupling of the electron and ion outflows. 

Noting that the electron-only events are detected in a turbulent plasma environment, \citet{stawarz2019} studied the observational properties of corresponding magnetosheath turbulence. They found that these properties are largely consistent with the energy spectra and intermittency observed in other cases of sub-ion-scale plasma turbulence. They also found the instances of energy-spectrum steepening at the scales associated with the electron inertial length ($\lesssim 4d_e$), possibly indicating energy dissipation associated with the electron reconnection events. 

These recent observations may suggest that the electron-only reconnection events may, in fact, be a characteristic property of sub-proton turbulence itself. This question is studied in the present work. A general association between turbulence and magnetic reconnection has long been proposed. More recently, it has received solid backing from kinetic simulations~\citep[e.g.][]{Karimabadi2013} and theoretical analyses in both magnetohydrodynamic and kinetic regimes \cite{loureiro2017,mallet2017,mallet2017a,boldyrev_2017,loureiro2017a,loureiro2018,boldyrev2019,loureiro2019}. The latter works in particular suggest that magnetized turbulence may always contain a small-scale subrange where the energy cascade is mediated by the tearing instability.

In order to investigate whether such a subproton-scale energy cascade may indeed be the cause for the electron-only reconnection events, we study the 2D PIC numerical simulations of collisionless subproton-scale turbulence for two regimes. The first regime is characterized by a  small electron beta, in particular $\beta_e\ll \beta_i \lesssim 1$. (For each particle species \(\alpha\) the plasma beta is related to the ratio of their gyroscale (\(\rho_\alpha\)) to their inertial scale (\(d_\alpha\)), \(\beta_\alpha=\rho_\alpha^2/d_\alpha^2\)). Such parameters are common for the Earth's magnetosheath and also expected to be relevant for the vicinity of the solar corona that will soon be studied with the Parker Solar Probe mission\footnote{Close to the solar corona, it is expected that $\beta_e\ll\beta_i<1$ (say, $\beta_e\sim0.01$, $\beta_i\sim0.1$). However, at scales smaller than $\rho_i$, we expect that our analysis will still be applicable.}. Subproton-scale turbulence in such a regime has been recently studied analytically in \cite{boldyrev2019}, which provides a useful guidance for interpreting our results. We will refer to turbulence regime analyzed here as kinetic-Alfv\'en, which is implied to mean fully nonlinear plasma turbulence in the phase space region where linear modes are kinetic-Alfv\'en waves. This region is characterized by high obliquity, $k_\perp \gg k_\|$, and low frequency, $\omega \ll kv_{Ti}$, of the fluctuations  \cite[e.g.,][]{mangeney06,alexandrova08c,boldyrev2015,chen2016,chen_boldyrev2017}. The second studied regime is characterised by a large electron beta, $\beta_e\sim \beta_i \lesssim 1$, typical of larger heliospheric distances. We compare the results obtained in this regime with the results for low electron beta. 

We observe that the electron-only reconnection structures are indeed generated by kinetic turbulence, in both considered regimes. The electron reconnection events and the corresponding  electron-Alfv\'en outflows that we find are rather similar to those observationally detected  by \citet{phan2018}. Our findings lend support to the idea that such electron-only regime of magnetic reconnection is an inherent property of turbulence. In order to further support this suggestion, we notice that the analytic studies of tearing-mediated kinetic-Alfv\'en turbulence \cite[][]{boldyrev2019} indicate that subproton-scale turbulence imposes certain limitations on the aspect ratio of the generated electron current sheets. As a result, the current layers with dimensions less than the ion scales, required for the electron-only reconnection,  are naturally produced in such a turbulent flow.




\section{Simulations}
In this work, we identify and characterize reconnection events in subproton-scale plasma turbulence. We analyze the 2D particle-in-cell (PIC) simulations corresponding to the two regimes, $\beta_e=0.04$, $\beta_i=0.4$ and $\beta_e=\beta_i=0.5$. The simulations were conducted using the VPIC code \citep{bowers2008}. The ratio of the plasma electron frequency to the electron cyclotron frequency was \(\omega_{pe}/\Omega_{ce}=2\), and the particle mass ratio was  \(m_i/m_e=100\). The simulation plane was perpendicular to the mean magnetic field \(B_0\) oriented in the \textit{z} direction.  The simulation domain was square with sides of length \(L_x=L_y=8\pi d_i=80\pi d_e\approx251\,d_e\). The low electron beta simulation, corresponding to \(\beta_e=0.04\), \(\beta_i=0.4\), had the resolution of \(n_x=n_y=3456\) cells, 4000 particles per cell per species, and a time step satisfying \(\omega_{pe}\delta t\approx0.05\). The numerical setup for these simulations has previously been discussed in more detail in \cite{roytershteyn2018}. The other simulation, corresponding to \(\beta_e=\beta_i=0.5\), had the resolution of \(n_x=n_y=1024\) cells, 10000 particles per cell per species, and a time step \(\omega_{pe}\delta t\approx0.17\).

In both regimes, decaying turbulence was seeded by imposing randomly phased perturbations of the type 
\begin{eqnarray}
\delta\mathbf{B}=\sum_k\delta\mathbf{B}_k\cos(\mathbf{k}\cdot\mathbf{x}+\chi_k),\\
\delta\mathbf{V}=\sum_k\delta\mathbf{V}_k\cos(\mathbf{k}\cdot\mathbf{x}+\phi_k), 
\end{eqnarray}
with the wave numbers \(\mathbf{k}=\{2\pi m/L_x, 2\pi n/L_y\}\), with \(m=-2,...,2\) and \(n=0,...,2\).   The initial energy of the magnetic and kinetic fluctuations was about ${\cal E}\approx 0.01 B_0^2$. The turbulence then evolved for several eddy turnover times (the eddy turnover time being approximately $30 \Omega_{ci}^{-1}$) before our analysis was performed.

In order to find the $X$-points, we note that in a 2D plane, magnetic field can be computed from a potential function:
\begin{eqnarray}
\mathbf{B}_\perp=\nabla\times(\psi\hat{\mathbf{z}})=\frac{\partial\psi}{\partial y}\hat{\mathbf{x}}-\frac{\partial\psi}{\partial x}\hat{\mathbf{y}}.\label{Bperp}
\end{eqnarray}
Since \(\mathbf{B}_\perp\cdot\nabla\psi=0\), the in-plane magnetic field is tangent to the contour lines of \(\psi\), so this function must have a saddle point wherever there is reconnection. 
To find \(\psi\), we start from the Ampère-Maxwell law. In the non-relativistic regime, the displacement current can be dropped, obtaining
\begin{equation}
\nabla\times\mathbf{B}_\perp=\mathbf{J_z}.
\end{equation}
Using (\ref{Bperp}) we get the following Poisson equation
\begin{equation}
\nabla^2\psi=-J_z.\label{Poisson}
\end{equation} 
The equation above was solved using a 2D Poisson solver based on Matlab's fast Fourier transform. 

In order to find the saddle points of the magnetic potential, the first and second derivatives of $\psi$ were computed numerically. Since the numerical gradient of the potential is never exactly zero, we identified as saddle points those points where the determinant of the Hessian matrix was negative and the first derivatives were less than 1\% of the standard deviation of the derivatives themselves over the whole domain. Similar approach was used in several prior studies~\citep[e.g.][and references therein]{Haggerty2017}.

The data used was time averaged over the time interval corresponding to several consecutive time steps (approximately spanning $1/5$th of the electron gyration period), which considerably reduced noise in the out-of-plane electric field.  Additional Fourier and Gaussian space filtering~\citep{Haggerty2017} had a very weak effect on the results, likely because the statistical noise was already low thanks to the large number of particles and time averages. Consequently, only the time average was used to obtain the results presented here.  

\begin{figure*}[tbh!]
\includegraphics[width=\columnwidth,height=0.9\columnwidth]{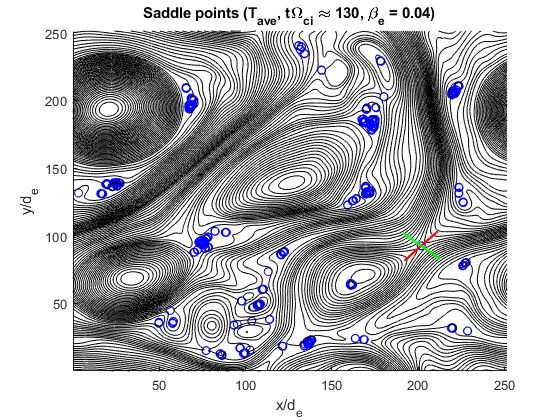}
\includegraphics[width=\columnwidth,height=0.9\columnwidth]{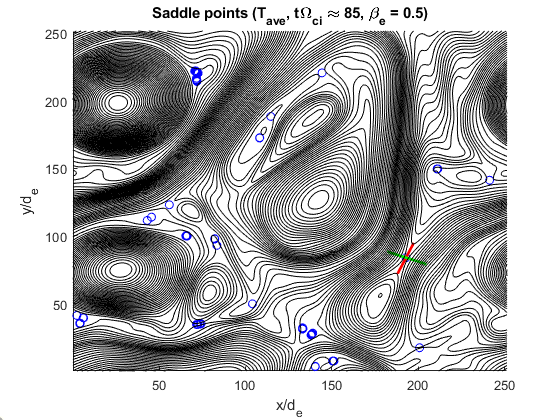}
\caption{Contour lines of the magnetic potential $\psi$. The left panel corresponds to kinetic-Alfv\'en turbulence with $\beta_e=0.04$, the right one to $\beta_e=0.5$. Saddle points of  $\psi$ are denoted by blue circles.  The electron-only reconnection events are discovered in both cases; the specific regions analyzed in detail below are marked by the green-red $X$ signs.  The directions of the red and green lines mark the (orthogonal) directions of the Hessian eigenvectors at the corresponding saddle points. \label{recevent} }
\end{figure*}
\begin{figure*}[!htb]
\includegraphics[width=\columnwidth]{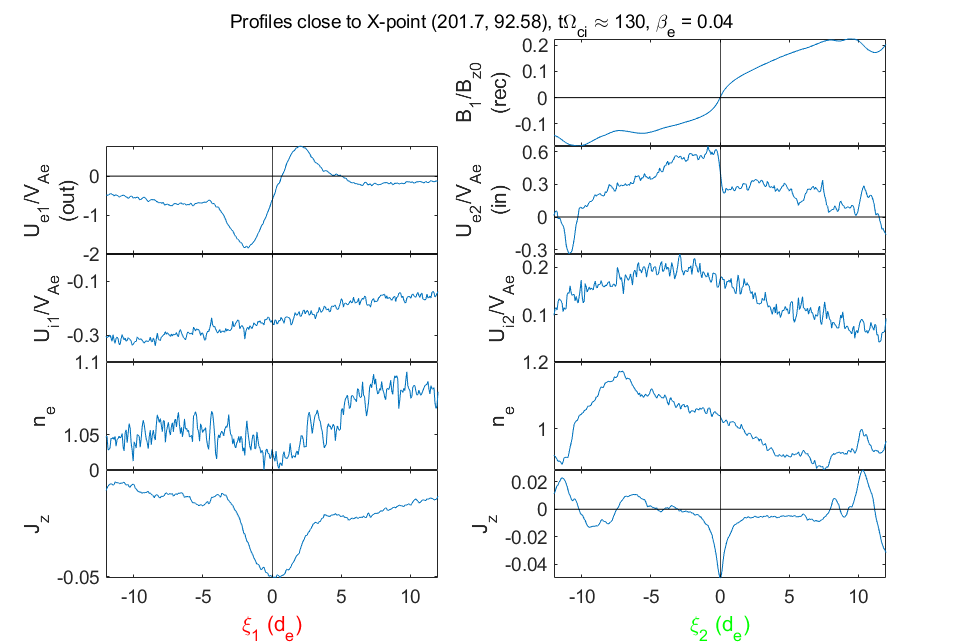}
\includegraphics[width=\columnwidth]{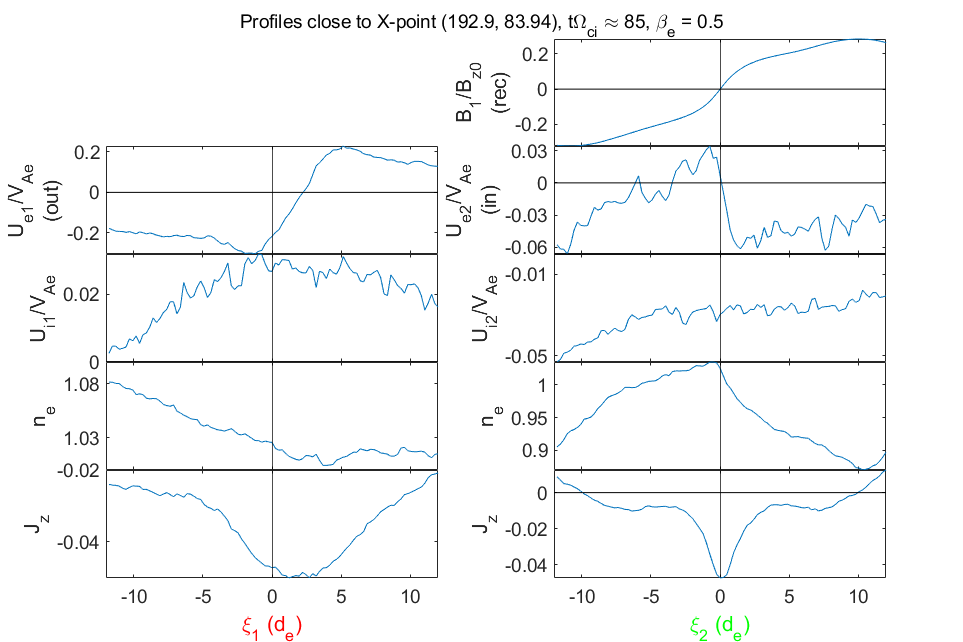}
\caption{Profiles close to the $X$-points marked in Figure~\ref{recevent}. Direction \(\xi_1\) corresponds to the Hessian eigenvector associated with the outflow direction, while \(\xi_2\) is close to the inflow direction. The $X$-point found by the Matlab algorithm is located at \(\xi_1=\xi_2=0\). In the \(\beta_e=0.5\) example, the peak of the \(J_z\) profile and the zero of the outflow velocity deviate from the position of the $X$-point, which is expected since the reconnection layer is not symmetric \cite[e.g.,][]{cassak2007,doss2015}. In both cases, there also appears to be a large-scale plasma flow at the X-point, which formally leads to nonzero mean electron and ion flows in the reconnection layer. \label{profiles} }
\end{figure*}

\section{Results}
\label{recrate}
Figure \ref{recevent} shows the saddle points found at some typical snapshots in \(\beta_e=0.04\) and \(\beta_e=0.5\) simulations. A reconnection site was marked in each case with an $X$, with the red line corresponding to the direction of the outflow and the green line to the inflow.  Figure \ref{profiles} shows the corresponding reconnecting component of the magnetic field (\(B_1\)), the electron inflow and outflow velocities (\(U_{e2}\) and \(U_{e1}\)), the ion inflow and outflow velocities (\(U_{i2}\) and $U_{i1}$), the electron density (\(n_e\)), and the out-of-plane current profile  (\(J_z\)) along the inflow and outflow directions. The velocities were normalized to the electron Alfv\'en velocity computed with the reconnecting magnetic field. The latter was estimated as half the jump of the magnetic field across the electron inflow layer. The magnetic field was normalized by the guide field \(B_0\), which is perpendicular to the simulation plane. Remarkably,  we find that in these reconnection sites, the ions do not couple to the electron inflow and outflow, which is similar to the observational results by \citet{phan2018}.

In order to understand why the ions do not couple to the electron motion, it is instructive to analyze the sizes of the out-of-plane current sheets characterized by the current density \(J_z\).   In the case $\beta_e=0.04$ (left panel in Figure~\ref{profiles}), the thickness of the current sheet is about $a\sim  d_e$, while its width is about $w\sim 4d_e$. In the case of $\beta_e=0.5$ (right panel in Figure~\ref{profiles}), the current sheet dimensions are slightly larger. In both cases, however, the sizes of the current sheets and of the electron outflow regions are significantly smaller than the typical scale ($\sim 10 d_i$), which has been proposed to be necessary for the electron outflows to couple to the ions  \cite[e.g.,][]{mandt1994,pyakurel2019,phan2018}. 

In addition to looking at the aforementioned current sheets we developed an algorithm to automatically identify and measure the dimensions of all out-of-plane current sheets potentially corresponding to reconnection events. To construct the current sheets we first identify all the points of the grid where the current is above a certain threshold (chosen to be 3 times the root mean square current). Then we find the local maxima of the current among those points. This is done by looking at the current on a square window of side \(2n+1\) centered on the point of interest (we used \(n=3\)). If the current at that point is the maximum in that window, it is regarded as a local maximum. Once the local maxima have been found they are ordered from highest to lowest. Starting on the point with the highest current, its four closest neighbours are checked, and if the current on any of them is found to be above a certain threshold (chosen to be some fraction of the peak corresponding to that current sheet) that point is taken to be part of the current sheet. The neighbors of the new points are then checked and so on until the current on all the points surrounding the current sheet is below the threshold (while for all the points in the current sheet it is above). When this process is completed for one maximum, the program moves on to the next local maximum. If this one belonged to any of the current sheets already found the program skips to the next maximum. The current sheets that overlap with the saddle points found previously are said to correspond to reconnection events. This algorithm, which was implemented in Matlab, is based on the algorithm previously discussed in \cite{zhdankin_statistical_2013}.

\begin{figure*}[tbh!]
\includegraphics[width=\columnwidth,height=0.9\columnwidth]{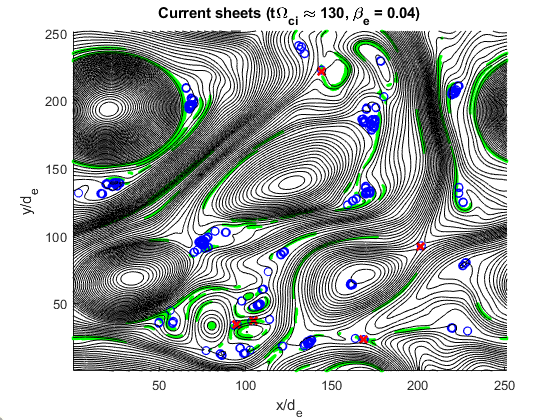}
\includegraphics[width=\columnwidth,height=0.9\columnwidth]{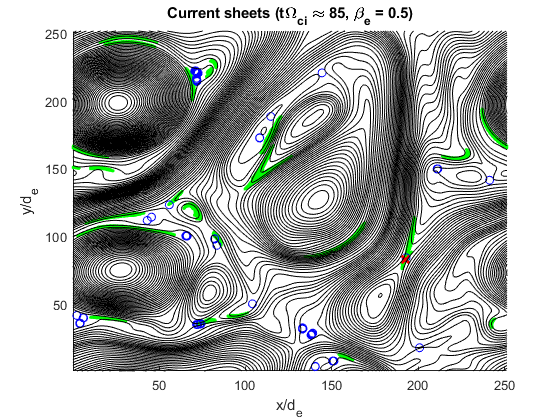}
\caption{Current sheets found (in green). The red Xs are the current peaks of those sheets containing saddle points (blue circles).}\label{Jsheets}
\end{figure*}

The width \(w\) of a current sheet is taken to be the largest distance between any two points belonging to the sheet. The thickness \(a\) is measured as the size of the current sheet in the direction of most rapid descent from the peak. Figure \ref{Jsheets} shows the current sheets found in both the low and high electron beta simulations, at a particular time, with the threshold that defines the current sheets set to 3/4 of the sheet's peak. The red X's highlight the current peaks corresponding to current sheets that overlap with saddle points (potential reconnection sites). Interestingly, in the high electron beta case the algorithm was able to identify only one current sheet containing a saddle point, which is the current sheet corresponding to the example discussed above.\footnote{In order to understand how generic such a situation is, we applied the  algorithm to several other randomly selected snapshots in both the low and high electron beta simulations, finding from 5 to 8 reconnection sites per snapshot in the former case and from 1 to 3 in the latter. This suggests that these electron-only reconnection events are easier to generate in the low electron beta environment.} The current sheet width determined by our algorithm was 14.6~\(d_e\) and the thickness was 1.23~\(d_e\). In the low electron beta case, five such sheets were found by our algorithm, including the example that we discussed previously in Figure~\ref{profiles}. The obtained current-sheet widths ranged from 3.2~\(d_e\) to 18.1~\(d_e\) (the average was 7.9~\(d_e\)). The thicknesses ranged from 0.36~\(d_e\) to 0.95~\(d_e\) (the average was 0.57~\(d_e\)). 

We checked that out of the five reconnecting current sheets found by the algorithm in the low electron beta case, the profiles corresponding to the X-points with coordinates \((x/d_e,y/d_e)=(104.4,37.38)\) and \((x/d_e,y/d_e)=(143.9,221.7)\) have a structure qualitatively similar to the electron-only reconnection event shown in Figure~\ref{profiles}. For the remaining two points, with coordinates \((x/d_e,y/d_e)=(94.68,34.18)\) and \((x/d_e,y/d_e)=(168.4,23.71)\), it is harder to clearly identify an electron inflow and outflow due to the strongly asymmetric structures of the current sheets, as can be seen in Figure~\ref{Jsheets}.

The small lengths of the detected current sheets is likely related to the fact that these current sheets are self-consistently generated by kinetic-scale turbulence. Note that \cite{boldyrev2019} proposed that the kinetic-Alfv\'en turbulent cascade in low electron beta, $\beta_e\ll \beta_i\lesssim 1$, 3D turbulence generates sheared magnetic structures that become tearing-unstable when their aspect ratio scales, depending on the assumed shape of the magnetic profile,  either as $a/w\sim d_e/a$ or as $a/w\sim (d_e/a)^{1/2}$, where $a$ is the thickness, $w$ is the width, and it is assumed that $a>d_e$. Such turbulence naturally generates current sheets with rather limited aspect ratio, so that the dimensions of the electron reconnection layers remain smaller than the ion scales. We thus expect the results presented here to be general.\footnote{This should be understood as a qualitative statement. Indeed, the theory of \cite{boldyrev2019} deals with the tearing instability, which is an initial stage of reconnection, while here we discuss fully nonlinear current sheets appearing after an X-point collapse. Moreover, the analytic consideration involves order of magnitude scaling estimates that are valid up to an (unknown) numerical factor. Note also that intermittent effects are not considered by the theory.} Further, our simulations suggest that similar conclusion holds for the regimes with $\beta_i \sim \beta_e \sim 1 $. 
We see that the observed aspect ratio of the current sheets in the case $\beta_e\sim \beta_i$ is qualitatively similar to that of the low electron  beta case (even though the number of detected electron-only reconnecting sites in the large-beta case is smaller). Analytically, the similar behaviour in the low and high electron beta regimes can be understood in the following way. In \cite{passot2017} it is shown that the finite Larmour radius corrections to the nonlinear equations governing the dynamics of the low electron beta plasma are on the order of $k_\perp^2\rho_e^2$; that needs to be added to terms proportional to $k_\perp^2 d_e^2$. When $\rho_e\sim d_e$, the finite Larmour radius corrections are on the order of the inertial terms, so while they can renormalize the numerical coefficients, they are not expected to qualitatively change the dynamics.\footnote{We also note that a similar analysis can be conducted in the regime $\beta_e\sim\beta_i\ll1$, see \cite{mallet2019}.}

  Finally, we measured the reconnection rate of the electron only reconnection sites in both the low and high electron beta simulations. In order to estimate the reconnection rate, several traditional methods can be used. First, one can define the reconnection rate by using the magnetic and electric fields on the scale of the electron layer, which is by definition a {\em local} reconnection rate \cite[e.g.,][]{cassak2017}. We however found that in the case of low electron beta, the electron inflow velocity $U_{e,2}$ does not agree with the formally calculated E-cross-B velocity, $E_z/B_2$. The latter, measured at distance $\pm a$ from the midplane of the reconnection layer, produces a  several times larger value than the electron inflow velocity directly measured from Figure~\ref{profiles}. This may be related to the presence of strong gradients of the electric field, gradients of pressure, and electron inertial effects inside the structure. A second way of measuring the reconnection rate is by finding the ratio of the electron inflow velocity to the electron outflow velocity for each reconnecting current sheet. This method was also found to be unreliable, giving in some cases reconnection rates above unity. This may be due to the difficulties in identifying the right points to measure the inflow and outflow velocities in the case of asymmetric current sheets. 
  
  On the other hand, from the relation \(U_{\text{inflow}}/U_{\text{outflow}}\sim a/w\), which is valid in a steady state, one can define the reconnection rate as the aspect ratio of a current sheet, \(a/w\), which can be reliably measured by our algorithm. We, therefore, use the aspect ratio of the reconnecting current sheets as a proxy for their reconnection rates. The average reconnection rate obtained in this way in the low electron beta case was 0.093, while in the high electron beta case it was 0.084. The aspect ratio \(a/w\) was also calculated for other values of the threshold used to define the current sheets, producing qualitatively similar results. Rather interestingly, the results are close to $0.1$, which demonstrates that the previously established result on collisionless magnetic reconnection \cite[e.g.,][]{birn2001,comisso2016,cassak2017} also applies to the novel electron-only reconnection regime.
 
 

\section{Conclusions}
\label{conclusions}
In this work, we addressed the possible origin of the electron-only reconnection events recently discovered in the Earth's magnetosheath observations \cite{phan2018}. In order to effectively decouple from the ions, the electron reconnection layers and the corresponding electron outflow jets, should be confined to sufficiently small regions, smaller than about $5-10\,d_i$, as estimated in numerical simulations \cite[e.g.,][]{mandt1994,pyakurel2019}. Our results support the conjecture that such electron-only reconnection events may be a direct consequence of subproton turbulence in general, and the kinetic-Alfv\'en turbulence in particular.  We analyzed the 2D PIC numerical simulations of kinetic-Alfv\'en turbulence in two regimes, corresponding  to low electron beta, $\beta_e\ll \beta_i\sim 1$, and high electron beta, $\beta_e\sim \beta_i\sim 1$. We have observed that the electron-only reconnection events are naturally generated in both cases (although, possibly, more efficiently in the low-beta case), with the corresponding electron and ion profiles quite similar to those observed in the Earth's magnetosheath.  The width of the electron current layers is on average $8d_e$ {$(\lesssim d_i)$} in the low-beta regime and seems to be larger in the high beta regime, but still significantly below the scale of about $5-10\, d_i$ required for the electron-ion coupling.  

A  theory of tearing-mediated kinetic-Alfv\'en turbulence recently proposed in \cite{boldyrev2019} predicts that such turbulence should generate tearing unstable magnetic profiles whose scales are smaller than the ion inertial scale. Therefore, we expect that electron-only reconnection should be a general feature of such regimes. Note that the theory of \cite{boldyrev2019} was developed for the simplified case of small electron beta, $\beta_e\ll \beta_i\sim 1$. {Our study found that electron-only reconnection is also characteristic of large electron beta, $\beta_e\sim \beta_i\sim 1$}. Moreover, we observed that the  reconnection rates in both cases are comparable to the well known ``standard'' value of 0.1. Our analysis of low electron beta kinetic-Alfv\'en turbulence will be especially relevant, among other astrophysical and space environments, for the vicinity of the solar corona that will soon be studied with the new NASA's Parker Solar Probe mission.

\acknowledgements
 {This research was partially supported by the Los Alamos National Laboratory (LANL) through its Center for Space and Earth Science (CSES). CSES is funded by LANL's Laboratory Directed Research and Development (LDRD) program under project number 20180475DR.}  The work of SB was partly supported by the NSF under grant no. NSF PHY-1707272, by NASA under grant no. NASA 80NSSC18K0646, and by DOE grant No. DE-SC0018266. VR was partly supported by NASA grant NNX15AR16G. Computational resources were provided by the NASA High-
End Computing Program through the NASA Advanced Supercomputing Division at Ames Research Center. PIC simulations were conducted as a part of the Blue Waters sustained-petascale computing project, which is supported by the National Science Foundation (awards OCI-0725070 and ACI-1238993) and the state of Illinois. Blue Waters is a joint effort of the University of Illinois at Urbana-Champaign and its National Center for Supercomputing Applications. Blue Waters allocation was provided by the National Science Foundation through PRAC award 1614664.




\bibliography{electron_refs}



\end{document}